\documentclass[]{article}

\usepackage{amsmath, amsthm, amscd, amsfonts, amssymb, graphicx, color}
\usepackage[backref,colorlinks=true]{hyperref}
\usepackage{graphtex}
\usepackage{hyperref}
\usepackage{url}
\newcommand{\squeezeup}{\vspace{-2.5mm}}
\newcommand{\pf}{\noindent \textit{Proof}:\ }
\def\bs{\blacksquare}
\hypersetup{colorlinks=true,linkcolor=blue,urlcolor=blue}

\begin{document}
	
	\title{HARMONIC CENTRALITY IN SOME GRAPH FAMILIES}
	\date{} 
	\maketitle
	
	\author{JOSE MARI E. ORTEGA$^1$, ROLITO G. EBALLE$^2$}
	\begin{flushleft}	
	$^1$ Computational Science and Technology Department, Philippine Science High School Southern Mindanao Campus, Davao City, Philippines. \\
	E-mail: josemari.ortega@smc.pshs.edu.ph\\ 
	$^2$Department of Mathematics, College of Arts and Sciences, Central Mindanao University, Musuan, Bukidnon, Philippines.
	\end{flushleft} 
	 
	\newtheorem{theorem}{Theorem}[section]
	\newtheorem{lemma}[theorem]{Lemma}
	\newtheorem{proposition}[theorem]{Proposition}
	\newtheorem{corollary}[theorem]{Corollary}
	\newtheorem{question}[theorem]{Question}
	
	\theoremstyle{definition}
	\newtheorem{definition}[theorem]{Definition}
	\newtheorem{algorithm}[theorem]{Algorithm}
	\newtheorem{conclusion}[theorem]{Conclusion}
	\newtheorem{problem}[theorem]{Problem}
	
	\theoremstyle{remark}
	\newtheorem{remark}[theorem]{Remark}
	\numberwithin{equation}{section}
	
	\begin{abstract}
	One of the more recent measures of centrality in social network analysis is the normalized harmonic centrality. A variant of the closeness centrality, harmonic centrality sums the inverse of the geodesic distances of each node to other nodes where it is 0 if there is no path from one node to another. It is then normalized by dividing it by $m-1$, where $m$ is the number of nodes of the graph. In this paper, we present notions regarding the harmonic centrality of some important classes of graphs. 
	\end{abstract}
	
	\noindent 
	{\let\thefootnote\relax\footnote{{2010 MSC: 05C12, 05C82, 91D30 
				Received November 15, 2021; Accepted January 15, 2022 \\
				Published in: Advances and Applications in Mathematical Sciences 21(5), 2581-2598	\\
				\url{https://doi.org/10.5281/zenodo.6396942} 		}}}

	\section{Introduction}
	\label{intro}
	In graph theory and social network analysis, the notion of centrality is based on the importance of a node in a graph. In 1978, Freeman \cite{3} expounded on the concept of centrality being an important attribute of social networks and its characteristics relate to other important properties and processes. Rodrigues \cite{6}, however, discussed centrality as not having a formal definition and may not be unique. A street corner in an urban network may be considered central when it is the most accessed part of the map. While in a social network, a celebrity or politician is considered central because she can easily share information with her millions of followers with a simple click of a button. Since there is no agreed-upon definition of centrality, several measures have been proposed, each with its strengths and qualities. 
	
	The most common measures of centrality include degree centrality, closeness centrality, betweenness centrality, eigenvector centrality, and PageRank centrality. Wasserman and Faust \cite{2} categorized these measures according to whether they are for non-directional relations or  directional relations. Some authors classified them according to whether they are degree-based or shortest path-based types of centrality measures. 
	
	One of the more recent measures of centrality is Harmonic Centrality. Introduced in 2000 by Marchiori and Latora \cite{4}, it is a variant of closeness centrality. While closeness centrality of a vertex $u$ is defined as the reciprocal of the average length of the shortest path between $u$ and all other vertices in $G$, harmonic centrality reverses this and measures the sum of the reciprocals of the distances of $u$ from each vertex in $G$. It was invented to solve the problem of dealing with unconnected graphs by equating the reciprocal to 0 if there is no path from one node to another. The harmonic centrality is normalized by dividing it by $m-1$, where $m$ is the number of nodes in the graph. For related works with closeness and betweenness centrality of some graph families, see \cite{1}, \cite{5}, and \cite{7}. 
	
	In this paper, we present notions regarding the harmonic centrality of some important classes of graphs. 
	
	\section{Preliminaries}
	\label{sec:2}
	\begin{sloppypar}
		For formality, we provide some definitions of the main concepts discussed in this paper.
		

		\begin{definition} [Harmonic Centrality of a Graph] \label{def1}
			\normalfont \mbox{} \\
			Let $G = (V (G),E(G))$ be a nontrivial graph of order $m$. If $u \in V(G)$, then the harmonic centrality of vertex $u$ is given by the expression
			$$	\mathcal{H}_{G} (u) =  \frac{\mathcal{R}_{G}(u)}{m-1} $$   
			where $\mathcal{R}_{G}(u)= \sum\limits_{u\neq x}\dfrac{1}{d(u,x)}$ is the sum of the reciprocals of the shortest distances $d(u,x)$ in $G$ between vertices $u$ and $x$, for all $x \neq u$, with $\dfrac{1}{d(u, x)}=0$ in case there is no path from $u$ to $x$ in $G$. 
		\end{definition} 
		\squeezeup	
		\begin{center}
			\begin{figure}[!htb]
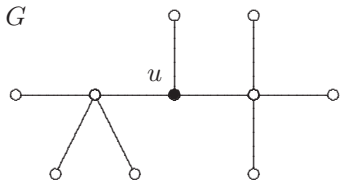

				\Magnify 1.5
				$$\pic 
				\Path (-200,20) (-180,20) (-160,20) (-140, 20) (-120, 20)
				\Path (-190,0) (-180,20) (-170,0)
				\Path (-160,20) (-160,40)
				\Path (-140,40) (-140,20) (-140,0)
				\Align [c] ($\bullet$) (-160,20)
				\Align [c] ($u$) (-165,25)
				\Align [c] ($G$) (-200,40)
				\cip$$
				\squeezeup \squeezeup \squeezeup
				\caption{A caterpillar graph $G$ with $u \in V(G)$, where $ \mathcal{H}_{G} (u) = \cfrac{2}{3}$}
				\label{fig:1} 
			\end{figure} 
		\end{center} 
		\squeezeup \squeezeup \squeezeup
		
		\begin{definition} [Harmonic Number $H_n$] \label{def2}
			\normalfont \mbox{} \\
			The $n$-th harmonic number  $H_n$ is the sum of the reciprocals of the first $n$ natural numbers, that is
			$$H_n = 1 +\cfrac{1}{2} +\cfrac{1}{3} + ... +\cfrac{1}{n} = \sum\limits_{k=1}^{n}\frac{1}{k}. $$ 
		\end{definition} 
		
		In this paper, the harmonic centrality of some special graphs such as path $P_m$, cycle $C_m$, fan $F_m$, wheel $W_m$, complete bipartite graph $K_{m, n}$, ladder $L_m$, crown $Cr_m$, prism $Y_m$, star $S_m$, book $B_m$, and helm graph $H_m$ are derived. Each graph considered is simple, finite, and undirected. 
		
\begin{center}
			\begin{figure}[!htb]
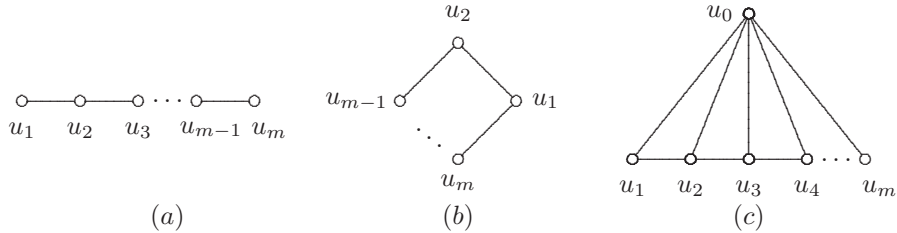

				\Magnify 1.1
				$$\pic 
				\Path (-200,20) (-180,20) (-160,20)
				\Path (-140,20) (-120,20)
				\Align[c] ($\ldots$) (-150,20)
				\Align [c] ($u_{1}$) (-200,10)
				\Align [c] ($u_{2}$) (-180,10)
				\Align [c] ($u_{3}$) (-160,10)
				\Align [c] ($u_{m-1}$) (-135,10)
				\Align [c] ($u_{m}$) (-115,10)
				\Path  (-70,20) (-50,40) (-30,20) (-50,0)
				\Align[c] ($\ddots$) (-60,10)
				\Align [c] ($u_{2}$) (-50,50)
				\Align [c] ($u_{m-1}$) (-85,20)
				\Align [c] ($u_{m}$) (-50,-8)
				\Align [c] ($u_{1}$) (-20,20)
				\Path (50,50) (50,0)
				\Path (50,50) (70,0)
				\Path (50,50) (90,0)
				\Path (50,50) (30,0)
				\Path (50,50) (10,0)
				\Path (10,0) (30,0)
				\Path (50,0) (30,0)
				\Path (70,0) (50,0)
				
				\Align [c] ($u_0$) (40,50)
				\Align [c] ($u_1$) (10,-10)
				\Align [c] ($u_2$) (30,-10)
				\Align [c] ($u_3$) (50,-10)
				\Align [c] ($u_4$) (70,-10)
				\Align [c] ($u_m$) (95,-10)
				\Align [c] ($\ldots$) (80,0)
				\Align [c] ($(a)$) (-150,-20)
				\Align [c] ($(b)$) (-50,-20)
				\Align [c] ($(c)$) (50,-20)
				\cip$$ \\[-40pt]
				\caption{$(a)$ Path $P_m$;\;$(b)$ Cycle $C_m$;\; and $(c)$ Fan $F_m$.}
				\label{fig:2}
			\end{figure}
		\end{center}

		The path $P_m$ of order $m$ is a graph with distinct vertices $a_1, a_2, ..., a_m$ and edges $a_{1}a_{2}, a_{2}a_{3}, ..., a_{m-1}a_m$. The cycle $C_m$ of order $m \geq 3$ is a graph with distinct vertices $a_1, a_2, ..., a_m$  and edges $a_{1}a_{2}, a_{2}a_{3}, ..., a_{m-1}a_m, a_{m}a_1$. The fan $F_m$ of order $m+1$, where $m\geq 3$, is formed by adjoining one vertex $u_0$ to each vertex of path $P_m = [u_1, u_2, ..., u_m]$. Figure \ref{fig:2} shows the skeletal graph for path, cycle, and fan graphs. 
		
\begin{center}
			\begin{figure}[!htb]
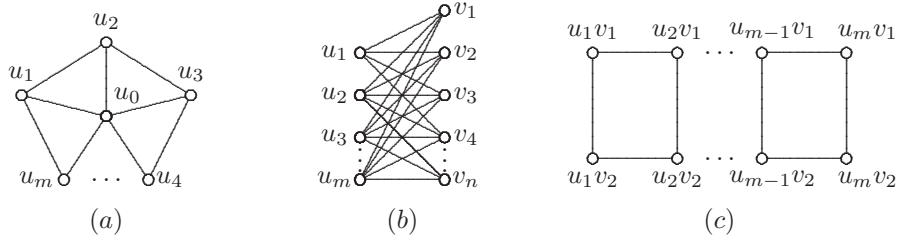

				\Magnify 0.8
				$$\pic 
				\Path (-70,40) (-90,80) (-50,105) (-10,80) (-30,40)
				\Path (-50,70) (-70,40)
				\Path (-50,70) (-90,80)
				\Path (-50,70) (-50,105)
				\Path (-50,70) (-10,80)
				\Path (-50,70) (-30,40)
				\Align[c] ($\ldots$) (-50,40)
				\Align [c] ($u_1$) (-90,90) 
				\Align [c] ($u_2$) (-50,115) 
				\Align [c] ($u_3$) (-10,90)
				\Align [c] ($u_4$) (-20,40)
				\Align [c] ($u_m$) (-83,40)
				\Align [c] ($u_0$) (-40,80) 
				
				\Path (70,100) (110,120)
				\Path (70,100) (110,100)
				\Path (70,100) (110,80)
				\Path (70,100) (110,60)
				\Path (70,80) (110,40)
				\Path (70,80) (110,120)
				\Path (70,80) (110,100)
				\Path (70,80) (110,80)
				\Path (70,80) (110,60)
				\Path (70,80) (110,40)
				\Path (70,60) (110,120)
				\Path (70,60) (110,100)
				\Path (70,60) (110,80)
				\Path (70,60) (110,60)
				\Path (70,60) (110,40)
				\Path (70,40) (110,120)
				\Path (70,40) (110,100)
				\Path (70,40) (110,80)
				\Path (70,40) (110,60)
				\Path (70,40) (110,40)
				\Align[c] ($\vdots$) (70,53)
				\Align[c] ($\vdots$) (110,53)
				\Align [c] ($u_{1}$) (58,100) 
				\Align [c] ($u_{2}$) (58,80) 
				\Align [c] ($u_{3}$) (58,60)
				\Align [c] ($u_{m}$) (58,40)
				\Align [c] ($v_{1}$) (120,120) 
				\Align [c] ($v_{2}$) (120,100) 
				\Align [c] ($v_{3}$) (120,80)
				\Align [c] ($v_{4}$) (120,60)
				\Align [c] ($v_{n}$) (120,40)
				\Path (180,100) (220,100)
				\Path (180,100) (180,50)
				\Path (180,50) (220,50)
				\Path (220,100) (220,50)
				\Path (260,100) (300,100)
				\Path (260,100) (260,50)
				\Path (260,50) (300,50)
				\Path (300,100) (300,50)
				\Align[c] ($\ldots$) (240,100)
				\Align[c] ($\ldots$) (240,50)
				\Align [c] ($u_{1}v_1$) (180,110) 
				\Align [c] ($u_{2}v_1$) (220,110)
				\Align [c] ($u_{m-1}v_1$) (265,110)
				\Align [c] ($u_{m}v_1$) (310,110)
				\Align [c] ($u_{1}v_2$) (180,40) 
				\Align [c] ($u_{2}v_2$) (220,40)
				\Align [c] ($u_{m-1}v_2$) (265,40)
				\Align [c] ($u_{m}v_2$) (310,40)
				\Align [c] ($(a)$) (-50,20)
				\Align [c] ($(b)$) (90,20)
				\Align [c] ($(c)$) (240,20)
				\cip$$ \\[-40pt]
				\caption{(a) Wheel graph $W_{m}$; (b) Complete bipartite graph $K_{m, n}$ ; and (c) Ladder $L_m$.}
				\label{fig:3}
			\end{figure}
		\end{center}

		The wheel graph $W_{m}$ of order $m+1$, $m>3$, formed by adjoining one vertex $u_0$ to each vertex of cycle $C_m = [u_1, u_2, ..., u_m]$.  The complete bipartite graph $K_{m, n}$  where both $m, n \geq 2$, $V(K_{m, n})=\{u_1, u_2, ..., u_m\} \bigcup \{v_1, v_2, ..., v_n\}$, $E(K_{m, n})=\{u_{i}v_{j}|1 \leq i \leq m, 1 \leq j \leq n\}$. The ladder $L_m$, of order $2m$, formed as the Cartesian product of a path graph $P_m = [u_1, u_2, ..., u_m]$ with the path graph $P_2=[v_1, v_2]$. Figure \ref{fig:3} shows the skeletal diagrams for a wheel, complete bipartite graph, and a ladder graph.
		
		\squeezeup 	\squeezeup 	\squeezeup 
		
		\squeezeup \squeezeup \squeezeup
		
\begin{center}
			\begin{figure}[!htb]
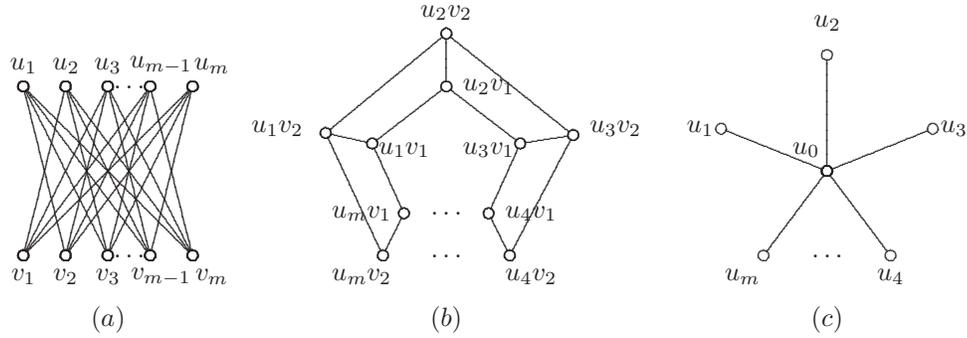

				\Magnify 0.8
				$$\pic 
				\Path (-120,140) (-100,60)
				\Path (-120,140) (-80,60)
				\Path (-120,140) (-60,60)
				\Path (-120,140) (-40,60)
				\Path (-100,140) (-120,60)
				\Path (-100,140) (-80,60)
				\Path (-100,140) (-60,60)
				\Path (-100,140) (-40,60)
				\Path (-80,140) (-100,60)
				\Path (-80,140) (-120,60)
				\Path (-80,140) (-60,60)
				\Path (-80,140) (-40,60)
				\Path (-60,140) (-100,60)
				\Path (-60,140) (-80,60)
				\Path (-60,140) (-120,60)
				\Path (-60,140) (-40,60)
				\Path (-40,140) (-100,60)
				\Path (-40,140) (-80,60)
				\Path (-40,140) (-60,60)
				\Path (-40,140) (-120,60)
				\Align[c] ($\ldots$) (-70,140)
				\Align[c] ($\ldots$) (-70,60)
				\Align [c] ($u_{1}$) (-120,150) 
				\Align [c] ($u_{2}$) (-100,150) 
				\Align [c] ($u_{3}$) (-80,150)
				\Align [c] ($u_{m-1}$) (-55,150)
				\Align [c] ($u_{m}$) (-31,150)
				\Align [c] ($v_{1}$) (-120,50) 
				\Align [c] ($v_{2}$) (-100,50) 
				\Align [c] ($v_{3}$) (-80,50)
				\Align [c] ($v_{m-1}$) (-55,50)
				\Align [c] ($v_{m}$) (-31,50)
				\Path (60,80) (45,113) (80,140) (115,113) (100,80)
				\Path (50,60) (23,118) (80,165) (140,117) (110,60)
				\Path (60,80) (50,60)
				\Path (45,113) (23,118)
				\Path (80,140) (80,165)
				\Path (115,113) (140,117)
				\Path (100,80) (110,60)
				\Align[c] ($\ldots$) (80,80)
				\Align[c] ($\ldots$) (80,60)
				\Align [c] ($u_{1}v_1$) (60,110) 
				\Align [c] ($u_{2}v_1$) (100,140) 
				\Align [c] ($u_{3}v_1$) (100,110)
				\Align [c] ($u_{4}v_1$) (120,80)
				\Align [c] ($u_{m}v_1$) (40,80)
				\Align [c] ($u_{1}v_2$) (0,120)
				\Align [c] ($u_{2}v_2$) (80,175)
				\Align [c] ($u_{3}v_2$) (160,120)
				\Align [c] ($u_{4}v_2$) (120,50)
				\Align [c] ($u_{m}v_2$) (40,50)
				\Path (260,100) (230,60)
				\Path (260,100) (210,120)
				\Path (260,100) (260,155)
				\Path (260,100) (310,120)
				\Path (260,100) (290,60)
				\Align[c] ($\ldots$) (260,60)
				\Align [c] ($u_0$) (250,110)
				\Align [c] ($u_1$) (200,120)
				\Align [c] ($u_2$) (260,170)
				\Align [c] ($u_3$) (320,120)
				\Align [c] ($u_4$) (290,50)
				\Align [c] ($u_m$) (220,50)
				\Align [c] ($(a)$) (-80,30)
				\Align [c] ($(b)$) (80,30)
				\Align [c] ($(c)$) (260,30)			
				\cip$$	\\[-40pt]
				\caption{(a) Crown $Cr_m$; (b)  Prism graph $Y_m$ ; and (c) Star $S_m$.}
				\label{fig:4}		
			\end{figure}
		\end{center}

		\begin{sloppypar} The crown graph $Cr_m$ of order $2m$ with $V(Cr_m)=\{u_1, u_2, ..., u_m\} \bigcup \{v_1, v_2, ..., v_m\}$ and whose edges are formed by adjoining $u_i$ to $v_j$ whenever $i \neq j$. The prism graph $Y_m$, of order $2m$ with $m \geq 3$, formed as the Cartesian product of a cycle graph $C_m=[u_1, u_2, ..., u_m]$ with the path graph $P_2=[v_1,v_2]$. The star graph $S_{m}$, of order $m+1$, $m>1$, formed by adjoining $m$ isolated vertices $u_i$, $1 \leq i \leq m$, to a single vertex $u_0$. Figure \ref{fig:4} shows the skeletal diagrams for a crown, prism, and star graph.
			
		\end{sloppypar}	
		
		\squeezeup \squeezeup \squeezeup
		
		\begin{figure}[!htb]
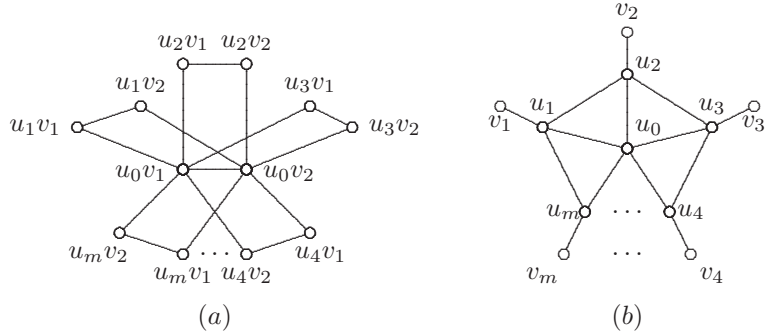

			\Magnify 0.8
			$$\pic 
			\Path (10,100) (-20,70)
			\Path (10,100) (-40,120)
			\Path (10,100) (10,150)
			\Path (10,100) (70,130)
			\Path (10,100) (40,60)
			\Path (40,100) (10,60)
			\Path (40,100) (-10,130)
			\Path (40,100) (40,150)
			\Path (40,100) (90,120)
			\Path (40,100) (70,70)
			\Path (10,100) (40,100)
			\Path (-20,70) (10,60)
			\Path (-40,120) (-10,130)
			\Path (10,150) (40,150)
			\Path (70,130) (90,120)
			\Path (40,60) (70,70)
			\Align[c] ($\ldots$) (25,60)
			\Align [c] ($u_{0}v_1$) (-10,98)
			\Align [c] ($u_{0}v_2$) (60,98)
			\Align [c] ($u_{1}v_1$) (-60,120)
			\Align [c] ($u_{1}v_2$) (-10,140)
			\Align [c] ($u_{2}v_1$) (10,160)
			\Align [c] ($u_{2}v_2$) (40,160)
			\Align [c] ($u_{3}v_1$) (70,140)
			\Align [c] ($u_{3}v_2$) (110,120)
			\Align [c] ($u_{4}v_1$) (75,60)
			\Align [c] ($u_{4}v_2$) (40,50)
			\Align [c] ($u_{m}v_1$) (10,50)
			\Align [c] ($u_{m}v_2$) (-30,60)
			\Path (200,80) (180,120) (220,145) (260,120) (240,80)
			\Path (220,110) (200,80)
			\Path (220,110) (180,120)
			\Path (220,110) (220,145)
			\Path (220,110) (260,120)
			\Path (220,110) (240,80)
			\Path (200,80) (190,60)
			\Path (180,120) (160,130)
			\Path (220,145) (220,165)
			\Path (260,120) (280,130)
			\Path (240,80) (250,60)
			\Align[c] ($\ldots$) (220,80)
			\Align[c] ($\ldots$) (220,60)
			\Align [c] ($u_{1}$) (180,130) 
			\Align [c] ($u_{2}$) (230,150) 
			\Align [c] ($u_{3}$) (260,130)
			\Align [c] ($u_{4}$) (250,80)
			\Align [c] ($u_{m}$) (190,80)
			\Align [c] ($u_0$) (230,120) 
			\Align [c] ($v_{1}$) (160,122)
			\Align [c] ($v_{2}$) (220,175)
			\Align [c] ($v_{3}$) (280,122)
			\Align [c] ($v_{4}$) (260,50)
			\Align [c] ($v_{m}$) (180,50)
			\Align [c] ($(a)$) (25,30)
			\Align [c] ($(b)$) (220,30)
			\cip$$ \\[-40pt]
			\caption{(a) Book $B_m$; and (b)  Helm graph $H_m$.}
			\label{fig:5}
		\end{figure}

		The book graph $B_m$ of order $2(m+1)$, formed as the Cartesian product of a star graph $S_m$ (with center vertex $u_0$) with path graph $P_2=[v_1, v_2]$. The helm graph $H_m$, $m \geq 3$, is obtained by adjoining a pendant vertex at each node of the $m$-ordered cycle of wheel $W_m$, with vertices $V(H_m)=[u_0, u_1, ..., u_m] \bigcup [v_1, v_2, ..., v_m]$. Figure \ref{fig:5} shows the skeletal diagrams for a book graph and a helm graph.
		
	\end{sloppypar}	
	
	\section{Main Results}
	\label{sec:3}
	
	In this section, we present some properties of harmonic centrality and their application to some graph families. Graphs considered in this paper are the path $P_m$, cycle $C_m$, fan $F_m$, wheel $W_m$, complete bipartite graph $K_{m, n}$, ladder $L_m$, crown $Cr_m$, prism $Y_m$, star $S_m$, book $B_m$, and helm graph $H_m$.
	
	\begin{theorem} \label{thm3.1} \normalfont Let $G$ be a nontrivial graph. Then each $u \in V(G)$ satisfies $0 \leq \mathcal{H}_{G}(u) \leq 1$.	
	\end{theorem} 
	
	\pf Let $u \in  V(G)$, then the vertex set $V(G)$ can be partitioned into three subsets, namely, (i)$N_G(u)$, (ii) $V(G)\setminus N[u]$, and (iii) the singleton vertex $u$.  
	
	(i) For the open neighborhood $N_G(u)$, 
	$$\mathcal{R}_{G}(u)\ = \sum\limits_{x \in N_G(u)} \frac {1}{\text{d}_G(u,x)} = \text{deg }u$$ 
	
	(ii) For $V(G)\setminus N[u]$, there will be two cases. Case 1, if $V(G)\setminus N[u]=\O$, then this case is the same with the first partition and we obtain
	$$  \mathcal{H}_{G}(u) = \frac{\mathcal{R}_{G}(u)}{m-1} = \frac {\text{deg }u}{m-1} =  \frac {m-1}{m-1} = 1 $$  
	
	For Case 2, if  $V(G)\setminus N[u]\neq \O$, then, 	
	$$\begin{aligned}[t]
		\mathcal{R}_{G}(u)\	& = \sum\limits_{x \in N_G(u)} \frac {1}{\text{d}_G(u,x)} + \sum\limits_{x \in V(G)\setminus N_G[u]} \frac {1}{\text{d}_G(u,x)} \\
		& < \text{deg}_G(u,x) + 1[m - 1 -\text{deg}_G(u)]  \\
		& = m-1
	\end{aligned}$$ 
	
	Thus, from this case we obtain
	$$  \mathcal{H}_{G}(u) = \frac{\mathcal{R}_{G}(u)}{m-1} < \frac {m-1}{m-1} = 1 $$  
	
	(iii) To complete graph $G$, we consider the singleton vertex $u$. Note that $\mathcal{H}_{G}(u) = 0$ whenever $u$ is not adjacent to any vertex in $G$. 
	
	Thus, combining these cases we get $0 \leq \mathcal{H}_{G}(u) \leq 1$, for all $u \in V(G)$. $\quad\bs$

	\begin{theorem} \label{thm3.2}  \normalfont Let $G$ be a nontrivial graph of order $m$ and let $u \in V(G)$. Then $\mathcal{H}_{G}(u)=1$ if and only if deg$_G(u)=m-1$.
	\end{theorem} 
	
	\pf If $\mathcal{H}_{G}(u)=1$, then $\cfrac{1}{m-1} \sum\limits_{x \in N_G[u]} \cfrac{1}{d(x,u)}=1$ and $\mathcal{R}_{G}(u)= \sum\limits_{x \in N_G[u]} \cfrac{1}{d(x,u)}=m-1$. This result can only happen whenever $\text{deg}(u)=m-1$, which means that vertex $u$ is adjacent and has a distance of 1 with all other  vertices in $G$.
	
	Assuming that $\text{deg}(u) \neq m-1$, then there exists at least one vertex $x$ of $G$ not adjacent to $u$. 
	For each $x \in V(G)\setminus N_G[u]$, we have $\frac{1}{d(x,u)} < 1$. 
	
	Now,
	$\begin{aligned}[t]
		\mathcal{R}_{G}(u)\	& = \sum\limits_{x \in N_G(u)} \frac {1}{\text{d}_G(u,x)} + \sum\limits_{x \in V(G)\setminus N_G[u]} \frac {1}{\text{d}_G(u,x)} \\
		& < \text{deg}_G(u,x) + 1[m - 1 -\text{deg}_G(u)]  \\
		& = m-1.
	\end{aligned}$  \\
	
	Thus, $\mathcal{R}_{G}(u)<m-1$ so that $\mathcal{H}_{G}(u)=\dfrac{\mathcal{R}_{G}(u)}{m-1} \neq 1. \quad \bs $

	\begin{corollary} \label{thm3.3}  \normalfont Let $G$ be a nontrivial connected graph of order $m$. Then $\mathcal{H}_{G}(u)=1$ for every $u \in V(G)$ if and only if $G=K_m$, where $K_m$ is the complete graph of order $m$. 
	\end{corollary} 
	
	\pf This follows from Theorem \ref{thm3.2} since the degree of each vertex of a complete graph is $m-1$. $\quad \bs$ \\
	
	\begin{theorem}  \label{thm3.4}  \normalfont For the path $P_m=[u_1, u_2, ..., u_m]$ of order $m \geq 2$, the harmonic centrality of any vertex $u_i, 1 \leq i \leq m$,  is given by
		\[
		\mathcal{H}_{P_m}(u_i) = \begin{cases}
			\frac{H_{m-1}}{m-1}  & \text{if $i=1$ or $i=m$}\\
			\frac{H_{i-1}+H_{m-i}}{m-1} & \text{if $1<i<m$.}
		\end{cases}
		\]
		\pf Consider a path $P_m=[u_1, u_2, ..., u_m]$ of order $m \geq 2$, then $\mathcal{R}_{P_m}(u_1) = \mathcal{R}_{P_m}(u_m) = 1 + \cfrac{1}{2} + ... +\cfrac{1}{m-1} = \sum\limits_{k=1}^{m-1} \frac{1}{k}=H_{m-1}$.  As for $u_i$, $1<i<m$, 	
		
		$\begin{aligned}[t]
			\mathcal{R}_{G}(u)\	& = \sum\limits_{j=1}^{m} \frac{1}{\text{d}_{P_{m}} (u_i, u_j)} \\
			& = \sum\limits_{j=1}^{i-1} \frac{1}{\text{d}_{P_{m}} (u_i, u_j)} + \sum\limits_{j=i+1}^{m} \frac{1}{\text{d}_{P_{m}} (u_i, u_j)}\\
			& = \Big[\frac{1}{i-1}+\frac{1}{i-2}+...+\frac{1}{3} +\frac{1}{2}+1\Big]+\Big[1+\frac{1}{2}+\frac{1}{3}+...+\frac{1}{m-i}\Big]\\
			& = \sum\limits_{j=1}^{i-1} \frac{1}{j} + \sum\limits_{j=1}^{m-i} \frac{1}{j}\\
			& = H_{i-1}+H_{m-1} 
		\end{aligned}$ \\
		Thus, \[
		\mathcal{H}_{P_m}(u_i) = \begin{cases}
			\frac{H_m}{m-1}  & \text{if $i=1$ or $i=m$}\\
			\frac{H_{i-1}+H_{m-i}}{m-1} & \text{if $1<i<m$.} \quad \bs
		\end{cases} 
		\] \\[-20pt]
	\end{theorem} 
	
	\begin{theorem} \label{thm3.5}  \normalfont For the cycle $C_m=[u_1, u_2, ..., u_m, u_1]$, the harmonic centrality of any vertex $u_i, 1 \leq i \leq m$ is given by
		\[
		\mathcal{H}_{C_m}(u_i) = \begin{cases}
			\frac{2}{m-1}\Big(H_{\frac{m-1}{2}}\Big)  & \text{if $m$ is odd}\\
			\frac{2}{m-1} \Big(H_\frac{m-2}{2}+\frac{1}{m}\Big) & \text{if $m$ is even.}
		\end{cases}
		\]
		
		\pf To prove this theorem, we need to consider the structure of a cycle $C_m$. For when $m$ is odd, a vertex $u$'s distances from each $x$ will be the same on both sides and $\mathcal{R}_{C_m}(u_i)$ can be computed as follows:
		$$\mathcal{R}_{C_m}(u_i) = \sum_{x \in V(C_m)} \frac{1}{d_{C_m}(u_i,x)} = 2\Big[1 +\frac{1}{2}+...+\frac{1}{\frac{m-1}{2}}\Big]=2H_\frac{m-1}{2}$$
		
		On the other hand, if $m$ is even, then a vertex directly opposite $u$ exists and $\mathcal{R}_{C_m}(u_i)$ can be computed as follows:
		$$\mathcal{R}_{C_m}(u_i) = \sum_{x \in V(C_m)} \frac{1}{d_{C_m}(u_i,x)} = 2\Big[1 +\frac{1}{2}+...+\frac{1}{\frac{m-2}{2}}\Big]+\frac{1}{\frac{m}{2}}=2H_\frac{m-2}{2}+\frac{2}{m}$$
		Normalizing these two cases form the harmonic centrality of cycle graph $C_m$. $\bs$
	\end{theorem} 
	
	\begin{theorem} \label{thm3.6}  \normalfont For the fan graph $F_m$ of order $m+1$, $m>2$, formed by adjoining one vertex $u_0$ to each vertex of path $P_m = [u_1, u_2, ..., u_m]$, the harmonic centrality of any vertex $u_i \in V(F_m)$ is given by
		
		\[
		\mathcal{H}_{F_m}(u_i) = \begin{cases}
			1 & \text{if $i=0$}\\
			\frac{m+2}{2m} & \text{if $i=1$ or $i=m$} \\
			\frac{m+3}{2m} & \text{if $1 < i < m$.}
		\end{cases}
		\]
		
		\pf Given that $u_0$ is adjacent to each vertex of path $P_m$, then by Theorem \ref{thm3.2}, $\mathcal{H}_{F_m}(u_0)=1$. For $u_1, u_m \in V(F_m)$, we have 
		
		$\begin{aligned}
			\mathcal{R}_{F_m}(u_1) = \mathcal{R}_{F_m}(u_m) \	& = \sum\limits_{x \in V(F_m)} \frac{1}{\text{d}_{F_{m}} (u_i, x)} \\
			& = \frac{1}{\text{d}(u_1, u_0)}+\frac{1}{\text{d} (u_1, u_2)}+\frac{1}{\text{d} (u_1, u_3)}+...+\frac{1}{\text{d} (u_1, u_m)}\\
			& = 1+1+\underbrace{\frac{1}{2}+...+\frac{1}{2}}_\text{$m-2$ addends} \\
			& = 2+\frac{1}{2}(m-2) \\
			& = \frac{m+2}{2} 
		\end{aligned}$ \\
		
		For $u_i \in V(F_m)$, if $1<i<m$, let us consider $u_2$ we have
		
		$\begin{aligned}
			\mathcal{R}_{F_m}(u_i) \	& = \sum\limits_{x \in V(F_m)} \frac{1}{\text{d}_{F_{m}} (u_i, x)} \\
			& = \frac{1}{\text{d}(u_2, u_0)}+\frac{1}{\text{d} (u_2, u_1)}+\frac{1}{\text{d} (u_2, u_3)}+...+\frac{1}{\text{d} (u_2, u_m)}\\
			& = 1+1+1+\underbrace{\frac{1}{2}+...+\frac{1}{2}}_\text{$m-3$ addends} \\
			& = 3+\frac{1}{2}(m-3) \\
			& = \frac{m+3}{2} 
		\end{aligned}$ \\
		
		To normalize, we divide $\mathcal{R}_{F_m}(u)$ by $m$ because fan graph $F_m$ has an order of $m+1$. $\bs$
	\end{theorem} 
	
	\begin{theorem} \label{thm3.7}  \normalfont For the wheel graph $W_{m}$ of order $m+1$, $m>3$, formed by adjoining one vertex $u_0$ to each vertex of cycle $C_{m}=[u_1, u_2, ..., u_m]$, the harmonic centrality of any vertex $u_i \in V(W_m)$ is given by
		
		\[
		\mathcal{H}_{W_m}(u_i) = \begin{cases}
			1 & \text{if $i=0$}\\
			\frac{m+3}{2m} & \text{if $1 \leq i \leq m$.}
		\end{cases}
		\]
		
		\pf Given that $u_0$ is adjacent to each vertex of cycle $C_m$, then by Theorem \ref{thm3.2}, $\mathcal{H}_{W_m}(u_0)=1$. For $u_i \in V(F_m),$ if $1 \leq i \leq m$, we have
		
		$\begin{aligned}[t]
			\mathcal{R}_{W_m}(u_i) \	& = \sum\limits_{x \in V(W_m)} \frac{1}{\text{d}_{W_{m}} (u_i, x)} \\
			& = \frac{1}{\text{d}(u_i, u_0)}+\frac{1}{\text{d} (u_i, u_{i-1})}+\frac{1}{\text{d} (u_i, u_{i+1})}+...+\frac{1}{\text{d} (u_i, u_m)}\\
			& = 1+1+1+\underbrace{\frac{1}{2}+...+\frac{1}{2}}_\text{$m-3$ addends}\\
			& = 3+\frac{1}{2}(m-3) \\
			& = \frac{m+3}{2} 
		\end{aligned}$ \\
		
		We normalize the wheel harmonic centrality by dividing it my $m$ since $W_m$ has $m+1$ vertices. $\bs$
	\end{theorem} 
	
	\begin{theorem} \label{thm3.8} \begin{sloppypar} \normalfont For the complete bipartite graph $K_{m, n}$ where both $m, n>0$, $V(K_{m, n})=\{u_1, u_2, ..., u_m\} \bigcup \{v_1, v_2, ..., v_n\}$, $E(K_{m, n})=\{u_i v_j|1 \leq i \leq m, 1 \leq j \leq n\}$, the harmonic centrality of vertices $u_i$ and $v_j$ are given by
			\[
			\mathcal {H}_{K_{m,n}}(u_i) = \frac {m+2n-1}{2(m+n-1)} 
			\text { and }
			\mathcal{H}_{K_{m,n}}(v_j) = \frac{2m+n-1}{2(m+n-1)}. 
			\]
			
			\pf Considering the specific structure of $K_{m,n}$ and the way the partite sets $\{u_1, u_2, ..., u_m\}$ and $\{v_1, v_2, ..., v_n\}$ are arranged, for $u_i \in V_1(K_{m,n}),$ we have
			
			$\begin{aligned}
				\mathcal{H}_{K_m}(u_i) \	& = \frac{1}{m+n-1} \Big( \sum\limits_{y \in V_2(K_{m,n})} \frac{1}{\text{d}_{K_{m,n}} (u_i, y)} +\sum\limits_{x \in V_1(K_{m,n})} \frac{1}{\text{d}_{K_{m,n}} (v_j, x)}\Big) \\
				& = \frac{1}{m+n-1} \Big(\big(m-1\big)\big(\frac{1}{2}\big)+ n\Big)\\
				& = \frac{m+2n-1}{2(m+n-1)} 
			\end{aligned}$ \\
			
			As for $v_j \in V_2(K_{m,n}),$ we have
			
			$\begin{aligned}
				\mathcal{H}_{K_m}(v_j) \	& = \frac{1}{m+n-1} \Big( \sum\limits_{y \in V_2(K_{m,n})} \frac{1}{\text{d}_{K_{m,n}} (u_i, y)} +\sum\limits_{x \in V_1(K_{m,n})} \frac{1}{\text{d}_{K_{m,n}} (v_j, x)}\Big) \\
				& = \frac{1}{m+n-1} \Big(m+ (n-1)\big(\frac{1}{2}\big)\Big)\\
				& = \frac{2m+n-1}{2(m+n-1)}. \quad \bs
			\end{aligned}$ \\
		\end{sloppypar} 
	\end{theorem} 
	
	\begin{theorem} \label{thm3.9} \begin{sloppypar} \normalfont For the ladder graph $L_m$ of order $2m$, formed as the Cartesian product of a path graph $P_m=[u_1, u_2, ..., u_m]$ with the path graph $P_2=[v_1, v_2]$, the harmonic centrality of any vertex $(u_i, v_j)$ is given by
			\[
			\mathcal{H}_{L_{m}}(u_i, v_j) =  \begin{cases}
				\frac{1}{2m-1} \Big(2H_{m-1}+\frac{1}{m}\Big) &  \text{for $i=1$ or $i=m, 1\leq j \leq 2$} \\
				\frac{1}{2m-1} \Big[2\Big(H_{i-1}+H_{m-i}\Big)+ \frac{1}{i} +\frac{1}{m-i+1}-1\Big]  & \text{for $1<i<m, 1\leq j \leq 2.$}   \\
			\end{cases}
			\]
			
			\pf Considering the structure of a ladder graph, we can group the vertices according the paths they belong to, that is, $[(u_1, v_1), (u_2, v_1),..., (u_m, v_1)] \in V_1(L_{m})$ and $[(u_1, v_2), (u_2, v_2),..., (u_m, v_2)] \in V_2(L_{m})$, so for $(u_1, v_j)$ and $(u_m, v_j)$ we have 
			
			\noindent 
			$\begin{aligned}
				\mathcal{R}_{L_m}(u_1,v_j) = \mathcal{R}_{L_m}(u_m,v_j) \	& = \sum\limits_{x \in V_1(L_{m})} \frac{1}{\text{d}_{L_{m}} ((u_1,v_1), x)} +\sum\limits_{x \in V_2(L_{m})} \frac{1}{\text{d}_{L_{m}} ((u_1,v_1), x)}  \\
				& = \frac{1}{\text{d}((u_1,v_1), (u_2,v_1))}+\frac{1}{\text{d} ((u_1,v_1), (u_3,v_1))}+...\\
				& \quad \quad ...+\frac{1}{\text{d} ((u_1,v_1), (u_{m-1},v_2))}+\frac{1}{\text{d} ((u_1,v_1), (u_m,v_2)}\\
				& = 1+\frac{1}{2}+...+\frac{1}{m-1}+ 1+\frac{1}{2}+...+\frac{1}{m-1}+\frac{1}{m} \\
				& = 2\sum_{k=1}^{m-1} \frac{1}{k} + \frac{1}{m} \\
				& = 2H_{m-1}+\frac{1}{m} 
			\end{aligned}$ \\
			
			As for $u_i$, $1<i<m$, 	
			
			\noindent
			$\begin{aligned}
				\mathcal{R}_{G}(u)\	& = \sum\limits_{x \in V_1(L_{m})} \frac{1}{\text{d}_{L_{m}} ((u_i,v_j), x)} +\sum\limits_{x \in V_2(L_{m})} \frac{1}{\text{d}_{L_{m}} ((u_i,v_j), x)}  \\
				& = \sum_{j=1}^{i-1} \frac{1}{\text{d}_{L_{m(V_1)}} (u_i, u_j)} + \sum_{j=i+1}^{m} \frac{1}{\text{d}_{L_{m(V_1)}} (u_i, u_j)}+
				\sum_{j=1}^{i-1} \frac{1}{\text{d}_{L_{m(V_2)}} (u_i, u_j)} + \sum_{j=i+1}^{m} \frac{1}{\text{d}_{L_{m(V_2)}} (u_i, u_j)}\\
				& = \big[\frac{1}{i-1}+\frac{1}{i-2}+...+\frac{1}{3} +\frac{1}{2}+1\big]+\big[1+\frac{1}{2}+\frac{1}{3}+...+\frac{1}{m-i}\big]\\
				& \quad \quad +\big[\frac{1}{i}+\frac{1}{i-1}+...+\frac{1}{3} +\frac{1}{2}\big]+\big[1+\frac{1}{2}+\frac{1}{3}+...+\frac{1}{m-i+1}\big]\\
				& = \big[\frac{1}{i-1}+\frac{1}{i-2}+...+\frac{1}{3} +\frac{1}{2}+1\big]+\big[1+\frac{1}{2}+\frac{1}{3}+...+\frac{1}{m-i}\big]\\
				& \quad \quad+\big[\frac{1}{i-1}+\frac{1}{i-2}+...+\frac{1}{3} +\frac{1}{2}+1\big]+\big[1+\frac{1}{2}+\frac{1}{3}+...+\frac{1}{m-i}\big] + \frac{1}{i}+\frac{1}{m-i+1}-1\\
				& = 2(H_{i-1}+H_{m-i}) + \frac{1}{i} +\frac{1}{m-i+1}-1
			\end{aligned}$ \\
			
			Thus, after normalizing we get
			\[
			\mathcal{H}_{L_{m}}(u_i, v_j) =  \begin{cases}
				\frac{1}{2m-1} \Big(2H_{m-1}+\frac{1}{m}\Big) &  \text{for $i=1$ or $i=m, 1\leq j \leq 2$} \\
				\frac{1}{2m-1} \Big[2\Big(H_{i-1}+H_{m-i}\Big)+ \frac{1}{i} +\frac{1}{m-i+1}-1\Big]  & \text{for $1<i<m, 1\leq j \leq 2.$}  \quad \bs \\
			\end{cases}
			\] \\[-20pt]
		\end{sloppypar} 	
	\end{theorem}
	
	\begin{theorem} \label{thm3.10} \begin{sloppypar} \normalfont For the crown graph $Cr_m$ of order $2m$ with $V(Cr_m)=\{u_1, u_2, ..., u_m\} \bigcup \{v_1, v_2, ..., v_m\}$ and whose edges are formed by adjoining $u_i$ to $v_j$ whenever $i \neq j$, the harmonic centrality of vertices $u_i$ and $v_j$ is given by
			\[
			\mathcal {H}_{Cr_m}(u_i) = \mathcal{H}_{Cr_m}(v_j) = \frac {9m-7}{12m-6}
			\]
			\pf Considering the structure of a crown graph $Cr_m$, the geodesic distance of $u_i$ to other $u$'s is 2. While distance between $u_i$ and $v_j$ is 3 whenever $i=j$. On the other hand, $u_i$ is adjacent to $v_j$, whenever $i \neq j$.
			
			Thus, 
			
			$\begin{aligned}
				\mathcal{H}_{Cr_{m}}(u_i) = \mathcal{H}_{Cr_{m}}(v_j) & = 
				\frac{1}{2m-1} \Big(\frac{1}{2}(m-1)+1(m-1)+ \frac{1}{3} \Big) \\
				& =\frac{1}{2m-1} \Big(\frac{9(m-1)+2}{6}\Big)  \\
				& = \frac{9m-7}{12m-6}.  \quad \quad \bs 
			\end{aligned}$
		\end{sloppypar}
	\end{theorem}

	\begin{theorem} \label{thm3.11} \normalfont For the prism graph $Y_m$, of order $2m$ and with $m \geq 3$, formed as the Cartesian product of a cycle graph $C_m=[u_1, u_2, .., u_m]$ with the path graph $P_2=[v_1, v_2]$, the harmonic centrality of any vertex $(u_i, v_j)$, is given by
		$$
		\mathcal{H}_{Y_m}(u_i,v_j) =  \begin{cases}
			\frac{1}{2m-1} \Big(4H_\frac{m-1}{2}-\frac{m-3}{m+1}\Big) & \text{if $m$ is odd}, 1 \leq j \leq 2  \\
			\frac{1}{2m-1} \Big(4H_{\frac{m}{2}}+\frac{2}{m+2}-\frac{m+2}{m}\Big)  & \text{if $m$ is even}, 1 \leq j \leq 2.
		\end{cases}
		$$
		
		\begin{sloppypar}		
			\pf Considering the construction of a prism $Y_m$ as a Cartesian product of a cycle graph $C_m=[u_1, u_2, .., u_m]$ and a path graph $P_2=[v_1, v_2]$, we can segregate the vertices according to the cycle they belong to. That is,  $[(u_1, v_1), (u_2, v_1),..., (u_m, v_1)] \in V_1(Y_{m})$ and $[(u_1, v_2), (u_2, v_2),..., (u_m, v_2)] \in V_2(Y_{m})$, so if $m$ is odd we have 
		\end{sloppypar}
		
		\noindent
		$\begin{aligned} 
			\mathcal{R}_{Y_m}(u_i,v_j)  \	& = \sum\limits_{x \in V_1(Y_{m})} \frac{1}{\text{d}_{Y_m} ((u_i,v_j), x)} +\sum\limits_{x \in V_2(Y_m)} \frac{1}{\text{d}_{Y_m} ((u_i,v_j), x)}  \\
			& = \Big[1+1+...+\frac{1}{\frac{m-1}{2}}+\frac{1}{\frac{m-1}{2}}\Big]+ \Big[1+\frac{1}{2}+\frac{1}{2}+...+\frac{1}{\frac{m+1}{2}}+\frac{1}{\frac{m+1}{2}}\Big] \\
			& = 4\sum_{k=1}^{\frac{m-1}{2}} \frac{1}{k} + 2\Big(\frac{2}{m+1}\Big) -1 \\
			& = 4H_{\frac{m-1}{2}}-\frac{m-3}{m+1} 
		\end{aligned}$ \\
		
		Now, if $m$ is even, we have
		
		\noindent
		$\begin{aligned} 
			\mathcal{R}_{Y_m}(u_i,v_j)  \	& = \sum\limits_{x \in V_1(Y_{m})} \frac{1}{\text{d}_{Y_m} ((u_i,v_j), x)} +\sum\limits_{x \in V_2(Y_m)} \frac{1}{\text{d}_{Y_m} ((u_i,v_j), x)}  \\
			& = \Big[1+1+\frac{1}{2}+\frac{1}{2}+...+\frac{1}{\frac{m}{2}}\Big]+ \Big[1+\frac{1}{2}+\frac{1}{2}+...+\frac{1}{\frac{m+2}{2}}\Big]\\
			& = 4\sum_{k=1}^{\frac{m}{2}} \frac{1}{k}  + \frac{1}{\frac{m+2}{2}} -1 -\frac{1}{\frac{m}{2}}  \\
			& = 4H_{\frac{m}{2}}+\frac{2}{m+2}-\frac{m+2}{m}\\ 
		\end{aligned}$ \\
		
		Normalizing and consolidating these results we get 
		$$
		\mathcal{H}_{Y_m}(u_i,v_j) =  \begin{cases}
			\frac{1}{2m-1} \Big(4H_\frac{m-1}{2}+\frac{3-m}{m+1}\Big) & \text{if $m$ is odd}, 1\leq j\leq2 \\
			\frac{1}{2m-1} \Big(4H_{\frac{m}{2}}+\frac{2}{m+2}-\frac{m+2}{m}\Big)  & \text{if $m$ is even}, 1\leq j\leq2. \quad \bs 
		\end{cases}
		$$ \\[-20pt] 
	\end{theorem}

	\begin{theorem} \label{thm3.12} \normalfont For the star graph $S_{m}$, of order $m+1$ with $m>1$, formed by adjoining $m$ isolated vertices $u_i, 1 \leq i \leq m$, to a single vertex $u_0$, the harmonic centrality of any vertex $u_i$ is given by
		\[
		\mathcal{H}_{S_{m}}(u_i) =   \begin{cases}
			1 & \text {for $i=0$ } \\
			\frac{m+1}{2m}  & \text{for $1 \leq i \leq m$.}
		\end{cases}
		\]
	\end{theorem} 
	
	\pf By theorem \ref{thm3.2},  $\mathcal{H}_{S_{m}}(u_0) = 1$ since all other vertices are adjacent to $u_0$. For $1 \leq i \leq m$, $u_i$ has a geodesic distance of 1 to $u_0$ and $\dfrac{1}{2}$, otherwise, thus we have
	
	$\begin{aligned} 
		\mathcal{H}_{S_m}(u_i)\	& = \frac{1}{m}\Big(1 + \frac{1}{2} (m-1)\Big) \\  	
		& = \frac{m+1}{2m}. \quad \bs   \\
	\end{aligned}$ \\ 
	
	\begin{theorem} \label{thm3.13} \normalfont For the book graph $B_m$ of order $2(m+1)$, formed as the Cartesian product of a star graph $S_m=[u_0, u_1,...,u_m]$ (with center vertex $u_0$ and order $m+1$) with path graph $P_2=[v_1,v_2]$, the harmonic centrality of any vertex $(u_i, v_j), 0 \leq i \leq m, 1\leq j\leq2$, is given by 
		\[
		\mathcal{H}_{B_m}(u_i) =  \begin{cases}
			\frac{3m+2}{4m+2} & \text{for $i=0, 1 \leq j \leq 2$ } \\
			\frac{5(m+2)}{6(2m+1)} & \text{for $1 \leq i \leq m, 1 \leq j \leq 2.$}\\
		\end{cases}
		\]
		
		\pf  For vertices $(u_0, v_1)$ and $(u_0, v_2)$, 
		
		$\begin{aligned} 
			\mathcal{H}_{B_m}(u_0, v_j)\	& = \frac{1}{2(m+1)-1}\Big[1(m+1) + \frac{1}{2} (m)\Big] \\  	
			& = \frac{3m+2}{4m+2} \\
		\end{aligned}$ \\ 
		
		For vertices $(u_i, v_j), 1 \leq i \leq m, 1 \leq j \leq 2$, 
		
		$\begin{aligned} 
			\mathcal{H}_{B_m}(u_i, v_j)\	& = \frac{1}{2m+1}\Big[2(1) + \frac{1}{2}(m) + \frac{1}{3}(m-1)\Big] \\
			& = \frac{1}{2m+1}\Big(\frac{12+3m+2m-2}{6}\Big) = \frac{5m+10}{12m+6}\\  	
			& = \frac{5(m+2)}{6(2m+1)}.  \quad \bs \\ 
		\end{aligned}$ \\ 
	\end{theorem} 
	
	\begin{theorem} \begin{sloppypar} \label{thm3.14} \normalfont For the helm graph $H_m$, of order $2m+1$ with $m \geq 3$, obtained by adjoining a pendant vertex to each node of the $m$-ordered cycle of wheel $W_m$, with vertices $V(H_m)=\{u_0, u_1, ..., u_m\} \bigcup \{v_1, v_2, ..., v_m\}$, the harmonic centrality of any vertex $u_i, 0 \leq i \leq m$ and $v_j, 1 \leq j \leq m$, is given by
			
			\[
			\mathcal{H}_{H_m}(u_i) =  \begin{cases}
				\frac{3}{4}  & \text{for $i=0$ } \\
				\frac{5m+15}{12m}  &  \text{for $1 \leq i \leq m$} \\
			\end{cases}
			\]
			\[
			\hspace{0.7em} \mathcal{H}_{H_m}(v_j) =  \begin{cases}
				\frac{7m+17}{24m} & \text{for $1 \leq j \leq m$. }
			\end{cases}
			\]
			\pf The distance of $u_0$ to the $m$-ordered $u_i$'s is 1, while its distance to the $m$-ordered $v_j$'s is $\cfrac{1}{2}$. So, $\mathcal{H}_{H_m}(u_0)=\cfrac{1}{2m}\Big(m(1)+m(\cfrac{1}{2})\Big)=\cfrac{3}{4}$. For $u_i, 1 \leq i \leq m$ 
			
			$\begin{aligned} 
				\mathcal{H}_{H_m}(u_i)\	& = \frac{1}{2m}\Big[4(1)+\frac{1}{2}(m-1) + \frac{1}{3} (m-3)\Big] \\  	
				& = \frac{5m+15}{12m} \\
			\end{aligned}$ \\ 
			
			As for $v_j, 1 \leq j \leq m$ , we have
			
			$\begin{aligned} 
				\mathcal{H}_{H_m}(v_j)\	& = \frac{1}{2m}\Big[1+(\frac{1}{2})3 + \frac{1}{3}(m-1) +\frac{1}{4}(m-3)\Big] \\  	
				& = \frac{7m+17}{24m}. \quad \bs \\
			\end{aligned}$ \\ 
		\end{sloppypar}
	\end{theorem} 
	
	\section{Conclusion}
	Harmonic centrality is a useful metric for analyzing graph structures. When compared to other centrality measures, harmonic centrality has the advantage of considering disconnected graphs. We have derived expressions for harmonic centrality of some graph families which are the basic components of larger and more complex networks. This study is therefore helpful for analyzing larger classes of graphs.


\begin{thebibliography}{widest-label}
		
		\bibitem{1} \href{https://doi.org/10.12988/ijcms.2021.91609} {R. Eballe, I. Cabahug, \textit{Closeness Centrality of Some Graph Families}, International Journal of Contemporary Mathematical Sciences, {\bf 14 (4)} (2021), 127--134.} 
		
		\bibitem{2} \href{https://doi.org/10.1017/CBO9780511815478} {K. Faust, S. Wasserman, \textit{Social Network Analysis: Methods and Applications}, Cambridge University Press, (1994).}
		
		\bibitem{3} \href{https://doi.org/10.1016/0378-8733(78)90021-7}{L.C. Freeman \textit{Centrality in social networks: Conceptual clarification},  Social Networks, {\bf 1 (3)} (1979), 215--239.}
		
		\bibitem{4} \href{http://dx.doi.org/10.1016/S0378-4371(00)00311-3} {M. Marchiori, V. Latora, \textit{Harmony in the small-world},  Physica A: Statistical Mechanics and Its Applications, {\bf 285 (3-4)} (2000), 539--546.}
		
		\bibitem{5} \href{https://doi.org/10.12988/ams.2021.914590} {M. Militante, R. Eballe, \textit{Weakly Connected 2-Domination in Some Special Graphs},  Applied Mathematical Sciences, {\bf 15 (12)} (2021), 579--586.}
		
		\bibitem{6} \href{https://arxiv.org/abs/1901.07901} {F. Rodrigues, \textit{Network centrality: an introduction},  A Mathematical Modeling Approach from Nonlinear Dynamics to Complex Systems, {\bf 15 (12)} (2018), n.pag.}
		
		\bibitem{7} \href{https://doi.org/10.1155/2014/241723} {S.K.R. Unnithan, B. Kannan, M. Jathavedan \textit{Betweenness Centrality in Some Classes of Graphs},  International Journal of Combinatorics, {\bf 2014} (2014), 12 pages.}
		
	\end{thebibliography}
\end{document}